\documentclass[12pt]{iopart}
\usepackage{iopams}  
\usepackage{graphicx}
\begin{document}

\newcommand{\Lam}{\ensuremath{\Lambda}}
\newcommand{\ALam}{\ensuremath{\bar{\Lambda}}}

\title[$\Lambda$ Production at Forward and Backward Rapidity]{$\Lambda$ Production at Forward and Backward Rapidity in d+Au Collisions at $\sqrt{s_{NN}} = 200$ GeV}

\author{F Simon (for the STAR Collaboration)\footnote{For the full author list and acknowledgments see appendix `Collaborations' of this volume.}} 

\address{Max--Planck--Institut f\"ur Physik, F\"ohringer Ring 6, 80805 M\"unchen, Germany}

\ead{fsimon@mppmu.mpg.de}

\begin{abstract}
Using the forward time projection chambers of the STAR experiment at the Brookhaven Relativistic Heavy Ion Collider we measure the centrality dependent \Lam\ and \ALam\ yields and the inverse slope parameters in d+Au collisions at forward and backward rapidities $y = \pm 2.75$. The contributions of different processes to stopping and baryon transport are probed exploiting the inherent asymmetry of the d+Au system. Comparisons to model calculations show that the baryon transport on the deuteron side is due to multiple collisions of the deuteron nucleons with gold participants. On the gold side, the inclusion of inital state nuclear effects or hadronic rescattering is necessary to describe the data.

\end{abstract}




\section{Introduction}

In relativistic heavy ion collisions at the highest RHIC energy, anti--baryon to baryon ratios close to unity are reached at mid--rapidity, pointing to an almost net--baryon free environment. Most particles originate from pair production. However, in contrast to the full transparency originally predicted \cite{Bjorken}, stopping over more than 5 units in rapidity is observed \cite{Brahms}. Away from mid--rapidity it has been found that in symmetric collisions contributions due to baryon transport from the projectile and target regions increase. These contributions start to dominate the particle yields at forward rapidity.

Due to their asymmetry, d+Au collisions offer a unique possibility to investigate the influence of different reaction mechanisms on stopping, ranging from baryon transport caused by multiple initial collisions of the participating nucleons to transport in later hadronic stages. In their passage through the gold nucleus, the nucleons of the deuteron suffer multiple collisions, while each participating gold nucleon experiences only one primary encounter.   

\section{Reconstruction of Forward \Lam}

The two radial--drift forward time projection chambers (FTPCs) of the STAR experiment permit the study of charged hadrons in the pseudorapidity range $2.5 < \vert \eta \vert < 4.0$ \cite{Ftpc}. The FTPCs cover both sides of the asymmetric d+Au collision. In general, full particle identification is impossible, only the sign of the charge can be determined. The momentum resolution is on the order of 15\%. Via the signature of delayed decays, however, the identification of long--lived particles becomes possible. 

The weakly decaying \Lam\ particles are reconstructed via their decay channel $\Lam \rightarrow p \pi$, which has a branching ratio of 64\%. Combinatorial background is eliminated with strict cuts on the vertex geometry. The remaining contribution, due to misidentified $K^0_S$, is obtained from a full detector simulation of HIJING \cite{Hijing} events in GEANT and is subtracted from the signal. 
\begin{figure}
	\begin{center}
		\includegraphics[width=1.0\textwidth]{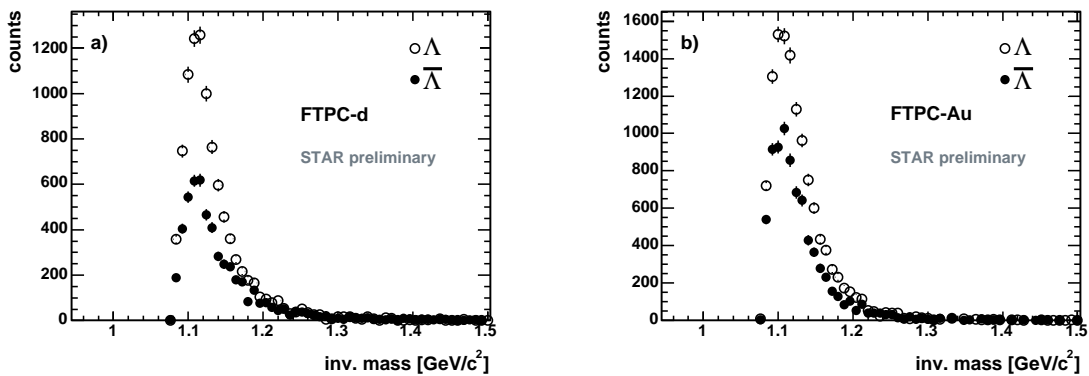}
		\vspace{-1cm} 
	\end{center}
	\caption{Invariant mass spectra for \Lam\ and \ALam\ in both detectors. The width of the peaks is due to the limited momentum resolution and is reproduced by simulations. The FTPC intercepting the outgoing Au fragments is labeled FTPC--Au, while the one on the deuteron side is refered to as FTPC--d.}
	\label{fig:MinvPlots}
\end{figure}
Figure \ref{fig:MinvPlots} shows the resulting invariant mass distributions of \Lam\ and \ALam\ in both detectors. The measured rapidity range is $|y| = 2.75 \pm 0.25$, with a transverse momentum acceptance of 0.5 GeV/c $<\,p_t\, <$ 2.0 GeV/c. 

\section{Particle Spectra and Centrality Dependence}

\begin{figure}
	\begin{center}
		\includegraphics[width=1.0\textwidth]{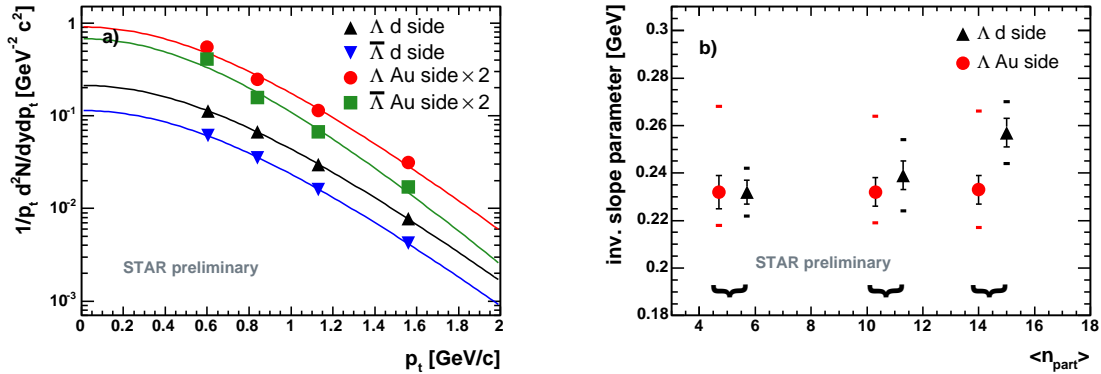}	
		\vspace{-1cm} 
	\end{center}
	\caption{a) Transverse momentum spectra of \Lam\ and \ALam\ for minimum bias events for $y = 2.75$ and $y = - 2.75$. Yields and inverse slope parameters are extracted with an $m_t$ exponential fit. b) Centrality dependence of the inverse slope parameter. While it is constant on the Au side, the parameter increases on the d side.}
	\label{fig:LambdaSpectraTemp}
\end{figure}

Using the track multiplicity at mid-rapidity to avoid auto correlations, the sample of 10.4 million minimum bias events ($<$n$_{part}\!>$ = 8.0) is subdivided into three centrality bins, the 20\% most central events ($<$n$_{part}\!>$ = 14.5), mid--central (20\% -- 40\%, $<$n$_{part}\!>$ = 10.8) and peripheral (40\% -- 100\%, $<$n$_{part}\!>$ = 5.1) events. The corresponding average numbers of participants for each centrality bin is obtained from HIJING. With efficiency and acceptance corrections determined by embedding simulated \Lam\ into real d+Au events, $p_t$ spectra are extracted for minimum bias and the three centrality classes. These spectra are corrected for feed down from $\Xi$ decays by using the measured mid-rapidity $\Xi / \Lam$ ratio. With an $m_t$ exponential fit to the data points, the inverse slope parameter and the total yield are determined. Figure \ref{fig:LambdaSpectraTemp}a shows the minimum bias transverse momentum spectra for \Lam\ and \ALam\ on both sides of the collision together with exponential fits in $m_t$. The transverse momentum range covered accounts for 65\% of the total yield.

The yield asymmetry in peripheral collisions is much less pronounced than in central collisions. On the gold side, the yield increases by a factor of 6 from peripheral to central events, while on the deuteron side there is only a 2.5 fold increase. 
The inverse slope parameter, shown in figure \ref{fig:LambdaSpectraTemp}b, exhibits no centrality dependence on the gold side, while there is a clear increase on the deuteron side with the number of participants. This is attributed to the increasing number of collisions in which the nucleons from the deuteron participate.

\section{Net \Lam\ Yield and \ALam/\Lam\ Ratio: Stopping and Baryon Transport}

Since the \Lam\ is a singly strange baryon and contains an $ud$ quark pair like the nucleon, both the net \Lam\ yield, i.e., \Lam\ - \ALam\ and the \ALam/\Lam\ ratio exhibit sensitivity to mechanisms of baryon transport with strip--off of one, two or all three quarks of the incoming baryon. By comparing the observations to results from a variety of models that use different processes for baryon transport, the contributions to stopping are investigated. 

The two HIJING models treat nucleus--nucleus collisions as a superposition of individual nucleon--nucleon interactions. In HIJING \cite{Hijing}, baryon transport originates from the initial reaction via mechanisms that lead to at least one initial valence quark in the final--state baryon. HIJING/B{$\overline{\mbox{B}}$} \cite{HijingB} invokes gluon junctions \cite{Junctions} to facilitate baryon transport independent from valence quark transfer. EPOS \cite{Epos} also treats only initial effects, but includes target and projectile excitation functions that describe the influence of target and projectile remnants on particle production. Among those considered, AMPT \cite{Ampt} is the only multi--phase model. In addition to HIJING--like initial processes it includes a hadronic transport stage.  

\begin{figure}
	\begin{center}
		\includegraphics[width=1.00\textwidth]{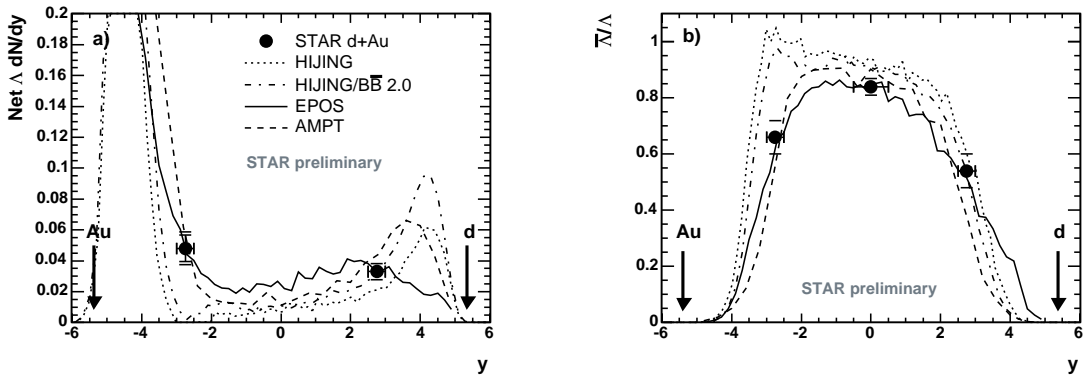}
		\vspace{-1cm} 
	\end{center}
	\caption{Net \Lam\ yield (a)) and \ALam/\Lam\ ratio (b)) for d+Au minimum bias events compared to model predictions. Significant baryon transport from both beam particles towards mid--rapidity is apparent. At mid--rapidity the analysis of the net \Lam\ yield  is still in progress while in b) the \ALam/\Lam\ ratio is given. Arrows mark beam rapidies.}
	\label{fig:ModelYields}
\end{figure}

Figure \ref{fig:ModelYields} shows the net \Lam\ yield and the \ALam/\Lam\ ratio together with the four model predictions. On the deuteron side ($y=+2.75$) all four models provide a fair description of the data while there is significant disagreement between the models at higher rapidity. On the gold side, only EPOS and AMPT do reasonably well. The nucleons from the deuteron experience several collisions as they pass through the gold nucleus, and no cold nuclear matter for final state interactions of the particles near deuteron rapidity is present. This scenario is well--described by the HIJING model. Additional baryon transport by gluon junctions in HIJING/B{$\overline{\mbox{B}}$} further improves the results. On the gold side, each participating nucleon only suffers one collision, and influences from the cold nuclear spectator matter become important. Only AMPT and EPOS which include initial nuclear contributions or final state interactions give a satisfactory description of the data.

\section{Conclusion}

With the forward TPCs in STAR, centrality dependent \Lam\ and \ALam\ yields and inverse slope parameters at $y = \pm 2.75$ have been measured in d+Au collisions at $\sqrt{s_{NN}} = 200$ GeV. The net \Lam\ yield and the \ALam/\Lam\ ratio show strong contributions from stopping. To reproduce the measurements on the gold side of the collision, models which include initial nuclear effects (EPOS) or a hadronic rescattering phase (AMPT) appear to be necessary. The mid-rapidity \Lam\ yield could decide between these two models.

\end{document}